\documentclass[11pt]{article}
\usepackage{moriond,epsfig}
\usepackage{latexsym,amssymb,psfrag}
\usepackage{graphicx}
\usepackage{pstricks}
\usepackage{amsmath}
\usepackage{subfigure}

\def\be{\begin{equation}}
\def\ee{\end{equation}}
\def\bea{\begin{eqnarray}}
\def\eea{\end{eqnarray}}

\begin{document}
\vspace*{4cm}
\title{GAUGINO NON-UNIVERSALITY AND NEUTRALINO DARK MATTER
\footnote{Invited talk given at the XXXVIIIth Rencontres de Moriond
session devoted to ELECTROWEAK INTERACTIONS AND UNIFIED THEORIES,
March 15-22, 2003, Les Arcs (France).}}

\author{ E. NEZRI }

\address{Service de Physique th\'eorique, Univesit\'e Libre de
Bruxelles.\\
Boulevard du triomphe, B-1050 Bruxelles, Belgium}

\maketitle\abstracts{ We study the effect of departures from SUSY GUT
  gaugino universality on
  the neutralino relic density, and both its direct and neutrino indirect
  detection. We find that the most interesting models are those with a
  value of $M_3|_{GUT}$ lower than the universal case.}

\section{Introduction - Universal case summary}

In a supersymmetric framework with $R-$parity conservation, the lightest
supersymmetric particle (LSP) is stable. In the Minimal Supersymmetric
Standard Model (MSSM), it is often the lightest neutralino ($\equiv$ {\it
  the} neutralino $\chi$) which is a neutral Majorana particle. It then
offers an interesting candidate to account for cold dark matter (CDM) in
the present Universe ($\Omega_{CDM}\sim 0.23$ recently confirmed by
WMAP results). For a review on neutralino dark matter and different detection
possibilities see \cite{Jungman:1996df}. Neutralino could be detected
by the energy they transfer to nuclei in direct detection
experiments (EDELWEISS \cite{Benoit:2002hf} \cite{EdelweissII}, ZEPLIN \cite{Zeplin}). Neutralinos can also be trapped
gravitationally in the centre of the Sun and annihilate giving rise to
muon neutrino fluxes detectable by neutrino telescopes such as ANTARES
\cite{AntarLee} or ICECUBE \cite{Ice3Edsjo}. We will study here both direct
detection and  neutrino indirect detection of neutralino captured in the Sun.

In ref. \cite{Myuniv} we studied the potential detection
of neutralino dark matter by neutrino telescopes and direct detection
experiments in CMSSM/mSugra models. Those models assume a
unification of the soft parameters of the MSSM at high energy $M_{GUT}\sim
2\times 10^{16}$ GeV reducing the 106 ``SUSY'' parameters of the MSSM down
to 5 : universal masses for scalars $(m_0)$ and gauginos $(m_{1/2})$,
universal trilinear $(A_0)$ and bilinear $(B_0)$ couplings, and a Higgs
``mass'' parameter $(\mu_0)$. Using renormalisation group equations (RGE)
and requiring radiative electroweak symmetry breaking, the usual input
parameters are $m_0$, $m_{1/2}$, $A_0$, $\tan{\beta}$ (the ratio of the 2
Higgs doublet vacuum expected values at low energy) and ${\rm
sign}(\mu)$. The neutralinos are the mass eigenstates coming from the
mixing of neutral gauge and Higgs boson superpartners. In the MSSM the
neutralino mass matrix in the
($\tilde{B},\tilde{W}^3,\tilde{H}^0_1,\tilde{H}^0_2$) basis is :

\begin{equation}
M_N=\left( \begin{array}{cccc}
M_1 & 0 & -m_Z\cos \beta \sin \theta_W^{} & m_Z\sin \beta \sin \theta_W^{}
\\
0 & M_2 & m_Z\cos \beta \cos \theta_W^{} & -m_Z\sin \beta \cos \theta_W^{}
\\
-m_Z\cos \beta \sin \theta_W^{} & m_Z\cos \beta \cos \theta_W^{} & 0 & -\mu
\\
m_Z\sin \beta \sin \theta_W^{} & -m_Z\sin \beta \cos \theta_W^{} & -\mu & 0
\end{array} \right)
\label{eq:matchi}
\end{equation}
and can be diagonalised by a single mixing matrix $z$ : $M_{\rm diag} = z M_N z^{-1}$.
The (lightest) neutralino is then given by the linear combination
\begin{equation}
\chi = z_{11}\tilde{B}     +z_{12} \tilde{W}^3
          +z_{13}\tilde{H}^0_1 +z_{14} \tilde{H}^0_2.
\end{equation}

\begin{figure}[t!]
\begin{center}
\begin{tabular}{cc}
\multicolumn{2}{c}{{ $ {\rm universal\ case}\ :\ A_0=0\ ;\ \tan{\beta}=45\ ;\ \mu>0$}}\\
\multicolumn{2}{c}{\includegraphics[width=\textwidth]{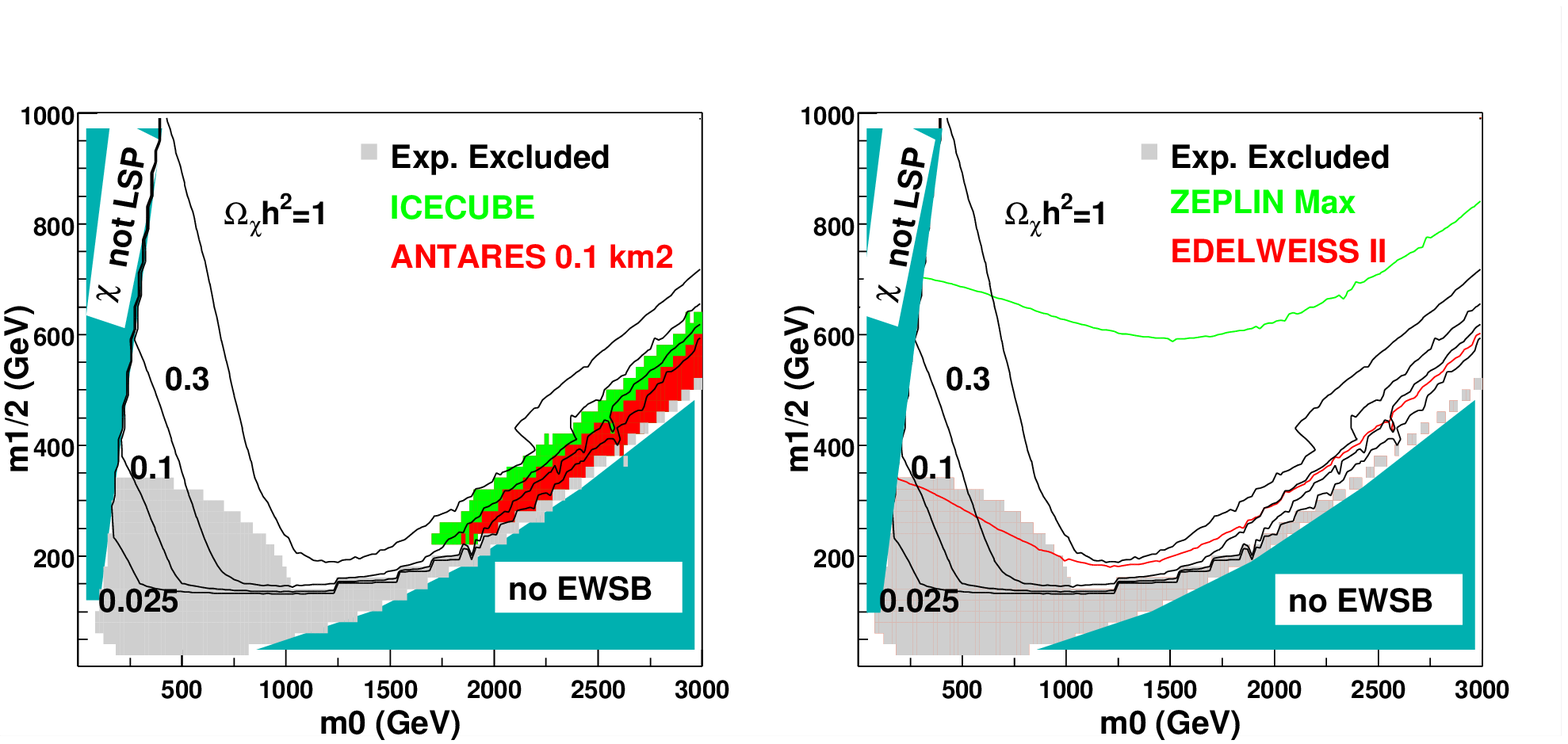}}\\
\hspace{0.25\textwidth}a) & \hspace{0.15\textwidth}b)
\end{tabular}
\caption{\small Neutralino relic density and detection potential in
the $(m_0,m_{1/2})$ plane for the universal case for a) neutrino
telescopes looking at muon fluxes coming from the Sun and b)
direct detection experiments.}
\label{fig:univ}
\end{center}
\end{figure}

We briefly summarised the CMSSM/mSugra universal situation shown on
figure \ref{fig:univ} \footnote{We used Suspect \cite{Djouadi:2002ze}
for RGE, potential minimisation and SUSY spectrum calculations and
Darksusy \cite{Darksusy} to estimate relic density and detection rates.} (Grey region are excluded by accelerator
constraints, see ref \cite{Mynonuniv} for details) :

\begin{itemize}
\item the low $m_0$ and $m_{1/2}$ region is strongly constrained by
  experimental limits on SUSY contributions to the $b\rightarrow s+\gamma$
  branching ratio, on the lightest Higgs mass and on SUSY contributions to
  $(g-2)_{\mu}$ : $a^{SUSY}_{\mu}$,
\item $\chi\tilde{\tau}$ coannihilation and  pseudo-scalar $A$ pole regions could
only be (not completly) detected  with very big future projects for
direct detection ($\sim$ 1 ton size),
\item the large $m_0$ ``focus point'' region \cite{Feng:1999zg,Fengindir} where the neutralino has a
  significant higgsino
  fraction ($f_H$) is interesting for direct and indirect detection
  experiments. Neutrino/muon fluxes coming from the Sun are large only in
  this region because of the $\chi\chi\xrightarrow{\chi^+_i,\chi_i}
  W^+W^-,\ ZZ$ and $\chi\chi\xrightarrow{Z}t\bar{t}$ channels which give
  rise to more energetic neutrino spectra and to muons with higher energy.
\end{itemize}

The Renormalisation Group Equations (RGE) lead SUSY models to a generic
hierarchy of particle spectrum in which scalars are heavier than light
neutralinos and charginos. Models with mixed neutralinos are indeed the only models leading to
large neutralino annihilation cross section $\sigma_{\chi-\chi}^A$
(important for its relic density and for indirect detection) {\it and} to
large neutralino--proton scalar and spin dependent elastic cross sections
(important for direct and indirect detection).

The main parameter of the MSSM through RGE is the gluino
mass parameter at GUT scale : $M_3|_{GUT}$ \cite{Kazakov:1999pe}. 
Thanks to $\alpha_s$, it
drives dominantly the evolution of squarks and Higgs soft masses and is
less dominant for sleptons.

In addition, a gaugino non-universality given by
$(M_2/M_1)_{GUT}<1$ can lead to a large wino component for the
neutralino and to an important modification of its couplings with
respect to the universal CMSSM case (where the neutralino is
essentially bino) and thus to a very
different phenomenology.

\section{Non-universality of gaugino soft masses at $M_{GUT}$}

We will now study the departure from universality of the two
parameters $M_2|_{GUT}$ and $M_3|_{GUT}$ and their benefits on the neutralino
relic density and the detection yields (See ref. \cite{Mynonuniv} for
the complete study and {\it e.g.} ref. \cite{Corsetti:2000yq}
\cite{birknel} for similar work concerning only direct detection). 

In the following, the departure from universality effects are
quantified by the ratios at $M_{GUT}$ $x=M_2/m_{1/2}$ or $x=M_3/m_{1/2}$ (with
$m_{1/2}=M_1=M_3$ or $m_{1/2}=M_1=M_2$ respectively) which will be
lowered starting from the CMSSM case ($x=1$).

\subsection{The $M_2|_{GUT}$ parameter}

\begin{figure}[t!]
\begin{center}
\begin{tabular}{ccc}
\multicolumn{3}{c}{{ $m_0=1500\ ;\ m_{1/2}=600\ ;\  A_0=0\ ;\ 
    \tan{\beta}=45\ ;\ \mu>0$}}\\
\hspace{2.8cm}a) & \hspace{4.5cm} b) & \hspace{1.5cm} c)\\
\multicolumn{3}{c}{\includegraphics[width=\textwidth]
  {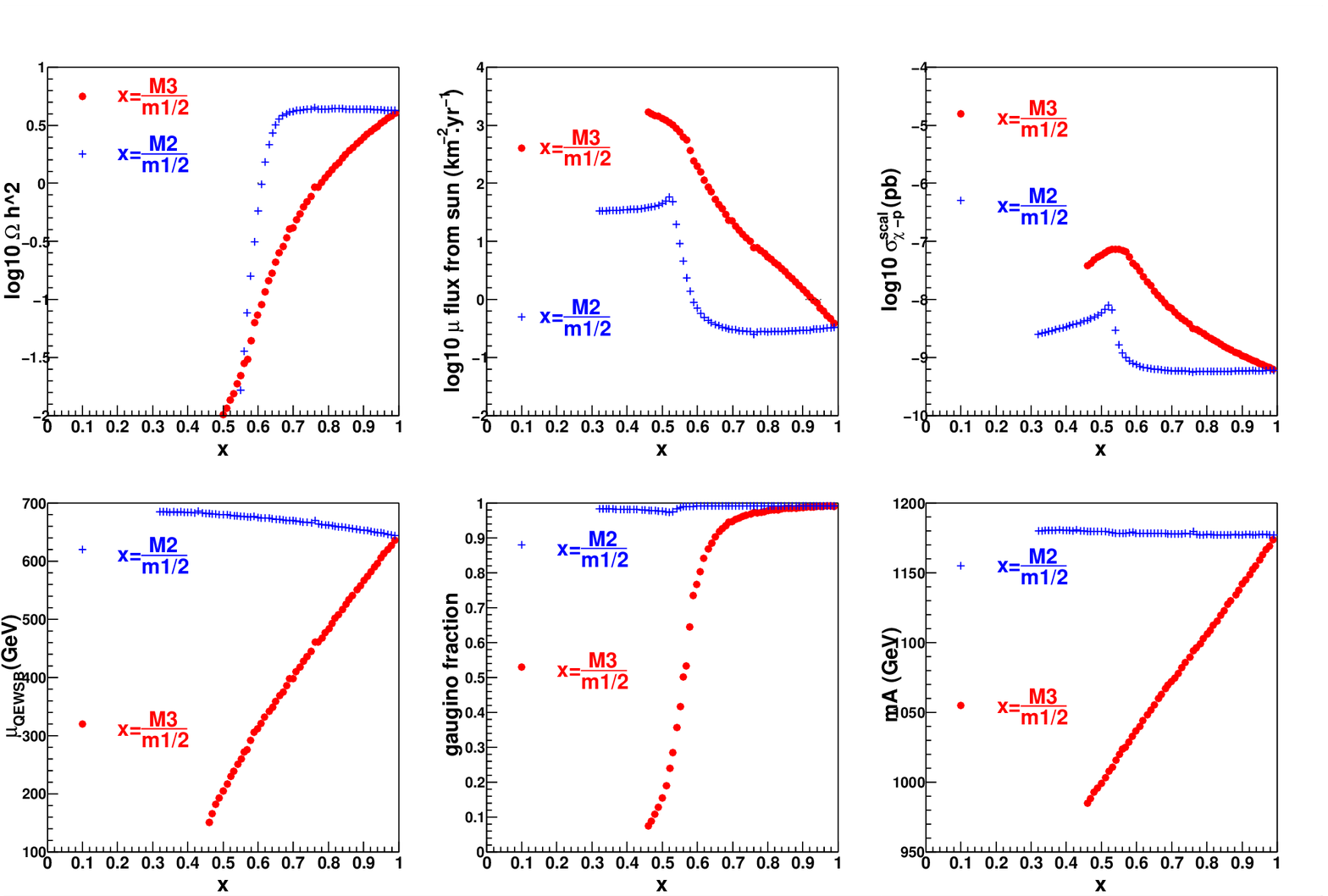}}\\
\hspace{2.8cm}d) & \hspace{4.5cm}e) & \hspace{1.5cm} f)
\end{tabular}
\caption{\small Evolution of a) the neutralino relic density, b) the
  muon flux coming from the Sun, c) the spin independent
  neutralino-proton cross section (direct detection), d) the $\mu$
  parameter, e) the gaugino fraction  and f) the pseudo-scalar mass
  $m_A$ (f) as functions of the $\frac{M_2}{m_{1/2}}$ and
  $\frac{M_3}{m_{1/2}}$ ratios for the CMSS Model with $m_0=1500$ {\rm
    GeV}, $m_{1/2}=600$ {\rm GeV}, $A_0=0$ {\rm GeV}, $\tan{\beta}=45$,
  $\mu>0$.}
\label{M3M2effect}
\end{center}
\end{figure}

The effect of the $M_2$ parameter is essentially a modification of the
neutralino composition. When the wino component of the neutralino
increases, the $\chi\chi\xrightarrow{\chi^+_1,\ \chi^0_2} W^+W^-,\ ZZ$
processes become more effective and enhance the annihilation cross section
$\sigma^A_{\chi-\chi}$ \cite{Birkwino}. In addition, the
strong $\chi\chi^0_2$ and $\chi\chi^+_1$ coannihilations become active and
the neutralino relic density strongly decreases. This wino component can
increase detection rates by an order of magnitude at most. Indeed, the
neutralino-quark coupling, which enters in direct detection
($\sigma^{scal}_{\chi-q}$, $H$ exchange) and in the capture for indirect
detection ($\sigma^{spin}_{\chi-q}$, $\tilde{q}$ exchange), is
$\tan\theta_W$-suppressed for pure bino w.r.t. pure wino.  The neutralino
annihilations into the hard $W^+W^-$ spectrum also give rise to more
energetic muons.  These enhancements of the (in)direct detection yields can
reach several order of magnitude with respect to the CMSSM case depending
on the values of the other SUSY parameters as can be seen for one
point in figures \ref{M3M2effect}b and \ref{M3M2effect}c.

However, the relevant value of $M_2$ is very critical, so its benefits are
only operative in a very narrow range. Given an $M_2/m_{1/2}\sim 0.6-0.7$
ratio (equivalent to $M_1|_{low}\sim M_2|_{low}$), the neutralino detection
yields are enhanced but the relic density drops down to much too small
values (see figure \ref{M3M2effect}a). In conclusion, the handling of this $M_2|{GUT}$ parameter
in order to get the desired neutralino dark matter phenomenology can only
be done by ``fine-tuning''. One way around this wino neutralino
extermination suggested in \cite{Moroi:1999zb} is to have the wino
neutralino population derived from AMSB models ($M_1|_{low}\simeq
3M_2|_{low}$) regenerated at low temperature by moduli decays which could
give a good relic abundance.

\subsection{The $M_3|_{GUT}$ parameter}

The impact of variations in the $M_3$ parameter is much more interesting.
It is indeed one of the key parameters of the MSSM through the RGE's
\cite{Kazakov:1999pe}. Its influence goes well beyond the neutralino sector. Indeed, a decrease of
$M_3|_{GUT}$ leads to a decrease of $m^2_{H_u}$ and of $\mu$ through the
radiative electroweak symmetry breaking mechanism, thus enhancing the
neutralino higgsino fraction, and leads also to a decrease of
$m_{\tilde{q}}$ and $m_A$. These effects are illustrated on the figures
\ref{M3M2effect}d,\ref{M3M2effect}e,\ref{M3M2effect}f. The neutralino
relic density then gradually decreases with $x=M_3/m_{1/2}$ (see figures
\ref{M3M2effect}a ) due to the
increase of the CMSSM dominant annihilation cross section channel (mainly
$\chi\chi\xrightarrow{A}b\bar{b}$, $\chi\chi\xrightarrow{Z}t\bar{t}$,
$\chi\chi\xrightarrow{\chi^+_i} W^+W^-$ and $\chi\chi\xrightarrow{\chi_i}
ZZ$ according to the CMSSM starting parameters \cite{Myuniv}). The
annihilation channels which directly depend on the neutralino higgsino
fraction finally dominate when $x$ is further lowered, because of the
decreasing of $\mu$.

For the CMSS Model with $m_0=1500$ {\rm GeV}, $m_{1/2}=600$ {\rm
GeV}, $A_0=0$ {\rm GeV}, $\tan{\beta}=45$, $\mu>0$ (figures
\ref{M3M2effect}), the dominant channel at
$x=1$ is $\chi\chi\xrightarrow{A}b\bar{b}$. By decreasing $x$, the
latter remains at first dominant while the neutralino relic density
is reduced due to the decrease of $m_A$ (figure \ref{M3M2effect}a and
f ), then since $\mu$ also decreases, the processes
$\chi\chi\xrightarrow{Z}t\bar{t}$ followed by
$\chi\chi\xrightarrow{\chi^+_i} W^+W^-$ and
$\chi\chi\xrightarrow{\chi_i} ZZ$ take successively over (as well as
the $\chi\chi^+$ and $\chi\chi^0_2$ coannihilations) further lowering
the relic density.

\noindent
{\bf Neutralino direct and indirect detections:}\\
When $x$ is lowered, the direct detection yields (see figures
\ref{M3M2effect}c ) can be
enhanced by 2 or 3 orders of magnitude with respect to the CMSSM case
($x=1$). Firstly the reduction of the squark masses favours the $\chi q
\xrightarrow{\tilde{q}} \chi q$ process, and mainly the coupling $C_{\chi q
  H}$ is maximal for maximum $z_{13(4)}z_{11(2)}$ mixed products which
increase with the neutralino higgsino fraction when $x$ decreases.
Moreover, $m_H\sim m_A$ and thus $\sigma^{scal}_{\chi-p}$ is enhanced due
to the decrease of $m_A$. When the neutralino gaugino fraction finally
drops, the $z_{13(4)}z_{11(2)}$ products decrease and
$\sigma^{scal}_{\chi-p}$ decreases back with $x$. This behaviour can
clearly be remarked on the figures \ref{M3M2effect}c, \ref{M3M2effect}e,
\ref{M3M2effect}f.

As far as neutrino indirect detection is concerned, the enhancement on the
muon fluxes coming from neutralino annihilation in the Sun, due to the
decrease of $M_3|_{GUT}$, can reach up to 6 orders of magnitude with
respect to the CMSSM case ($x=1$) \cite{Mynonuniv}. The main effect is coming from the
increase of the spin dependent neutralino-proton elastic cross section,
firstly due to the decrease of the squark masses which enhance $\chi
q\xrightarrow{\tilde{q}}\chi q$ in $\sigma^{spin}_{\chi-p}$, then mainly to
the decrease of $\mu$ leading to a larger higgsino fraction in the
neutralino which enhance $\chi q\xrightarrow{Z}\chi q$ in
$\sigma^{spin}_{\chi-p}$. Moreover the larger higgsino fraction also favours
the neutralino annihilations into the $\chi\chi\rightarrow W^+W^-,\ ZZ$ and
$\chi\chi\rightarrow t\bar{t}$ channels which give harder neutrino spectra
than $\chi\chi\rightarrow b\bar{b}$. This is illustrated on figures
\ref{M3M2effect}b, \ref{M3M2effect}d, \ref{M3M2effect}e. This
enhancement is not as peaked in $x$ as for direct detection but remains
maximum as long as the higgsino fraction dominates the neutralino
composition. However the relic density becomes very small when $x$ is
further lowered.

The picture of figure \ref{M3M2effect} is actually very generic. To summarise, it is possible, by lowering $M_3|_{GUT}$, to
decrease the neutralino relic density to the desired cosmological value for
any CMSS Model even for $m_0$ and $m_{1/2}$ as large as several TeV \cite{Mynonuniv}. The
value of $M_3|_{GUT}$ necessary to get a relic density $\Omega_{\chi} h^2
\sim 0.11$ mainly depends on $m_{1/2}$, a
quite generic $x$ value being:
\be
M_3|_{GUT}\sim 0.6(\pm0.1)m_{1/2}+{\rm corrections}(m_0,\tan{\beta},m_b)
\ee
These $x$ values
also favour the direct detection yields through the $\chi
q\xrightarrow{H}\chi q$ process. Moreover, the decrease of $M_3$ increases
the neutralino higgsino fraction and enlarges the ``focus-point''
corridor at large $m_0$ as well as the region of the parameter space
accessible to direct and indirect detection. 
\begin{figure}[t!]
\begin{center}
\begin{tabular}{cc}
\multicolumn{2}{c}{{ $A_0=0\ ;\ \tan{\beta}=45\ ;\ \mu>0\ ; M_3/m_{1/2}=0.63$}}\\
\multicolumn{2}{c}{\includegraphics[width=\textwidth]{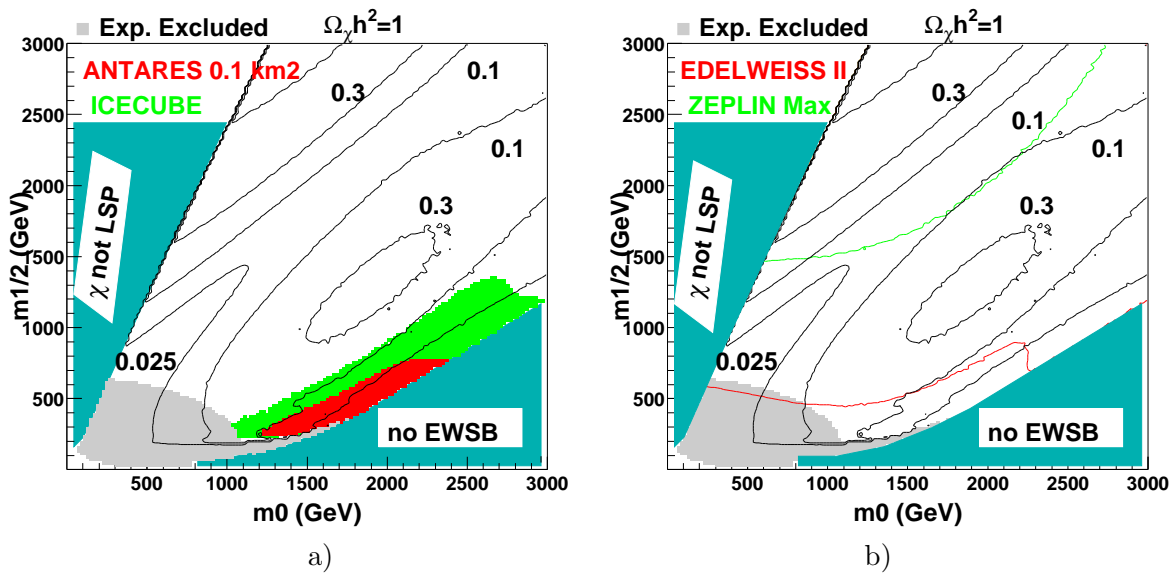}}\\
\hspace{0.25\textwidth}a) & \hspace{0.15\textwidth}b)
\end{tabular}
\caption{\small Neutralino detection potential in the $(m_0,m_{1/2})$
plane for $x=M_3/m_{1/2}=0.63$ for a) neutrino telescopes looking at
muon fluxes coming from the Sun and b)
direct detection experiments.}
\label{x063}
\end{center}
\end{figure}
An example with a fixed $M_3/m_{1/2}$
ratio is shown on figure \ref{x063} where regions with interesting relic
density and experiment sensitivity areas are vastly improved with respect
to the CMSSM case (figure \ref{fig:univ}), especially noticing the
much larger $m_{1/2}$ range shown.
\section{Conclusion}
In this paper, we have explored departures from CMSSM gaugino universality.
$M_2|_{GUT}$ and $M_3|_{GUT}$ respectively increase the
wino and the higgsino content of the neutralino when lowered away from
their universal values. The higgsino component is more efficient than the
wino one to improve the detection rates, making of $M_3|_{GUT}$ the most
relevant degree of freedom, as its value also affects the whole MSSM
spectrum. Such models with lower $M_3|_{GUT}$ values
have a better relic abundance and are much more promising from a detection
point of view, with rates increased by several orders of magnitude compared
to the universal case. 
\section*{Acknowledgments}
I'm grateful to Jean-Marie Fr\`ere and all the organizers for Moriond
invitation. I thank my collaborators on this work Vincent Bertin and
Jean Orloff.

\end{document}